\documentclass{iopart}
\eqnobysec
\newcommand{\be}{\[}
\newcommand{\ee}{\]}
\newcommand{\bea}{\begin{eqnarray*}}
\newcommand{\eea}{\end{eqnarray*}}

\newcommand{\beq}{\begin{equation}}
\newcommand{\eeq}{\end{equation}}
\newcommand{\beqa}{\begin{eqnarray}}
\newcommand{\eeqa}{\end{eqnarray}}
\newcommand{\bin}[2]{{#1\choose#2}}
\renewcommand{\d}{{\rm d}}
\newcommand{\dpar}{\partial}
\newcommand{\ds}[1]{{\displaystyle #1}}
\renewcommand{\e}{{\rm e}}
\newcommand{\eff}{{\rm eff}}
\newcommand{\ens}{{\left\{N_i\right\}}}
\newcommand{\enspr}{{\left\{N'_i\right\}}}
\newcommand{\eq}{{\rm eq}}
\newcommand{\fd}{fluc\-tu\-a\-tion-dis\-si\-pa\-tion }
\newcommand{\Fd}{Fluc\-tu\-a\-tion-dis\-si\-pa\-tion }
\newcommand{\frad}[2]{\displaystyle{\displaystyle#1\over\displaystyle#2}}
\newcommand{\g}{\gamma}
\renewcommand{\H}{{\cal H}}
\renewcommand{\i}{{\rm i}}
\renewcommand{\l}{\ell}
\renewcommand{\L}{\Lambda}
\newcommand{\lam}{\lambda}
\newcommand{\Leq}{{\Lambda_\eq}}
\newcommand{\mean}[1]{\left\langle#1\right\rangle}
\newcommand{\PP}{{\cal P}}
\newcommand{\sg}{{\rm sg}}
\newcommand{\X}{{\cal X}}

\begin{document}

\title{Nonequilibrium dynamics of urn models}

\author{C Godr\`eche\dag\footnote{godreche@spec.saclay.cea.fr}
and J M Luck\P\footnote{luck@spht.saclay.cea.fr}}

\address{\dag\ Service de Physique de l'\'Etat Condens\'e,
CEA Saclay, 91191 Gif-sur-Yvette cedex, France}

\address{\P\ Service de Physique Th\'eorique\footnote{URA 2306 of CNRS},
CEA Saclay, 91191 Gif-sur-Yvette cedex, France}

\begin{abstract}
Dynamical urn models, such as the Ehrenfest model,
have played an important role in the early days of statistical mechanics.
Dynamical many-urn models generalize the former models in two respects:
the number of urns is macroscopic, and thermal effects are included.
These many-urn models are exactly solvable in the mean-field geometry.
They allow analytical investigations of
the characteristic features of nonequilibrium dynamics referred to as aging,
including the scaling of correlation and response functions
in the two-time plane and the violation of the \fd theorem.
This review paper contains a general presentation of these models,
as well as a more detailed description of two dynamical urn models,
the backgammon model and the zeta urn model.
\end{abstract}

\section*{Prologue}

Urns containing balls, just as dice or playing cards,
are ubiquitous in writings on probability theory,
reminding us that this branch of mathematics owes its early developments
to practical questions arising in playing games.
Dynamical urn models, such as the Ehrenfest model,
have played an important role in the elucidation of conceptual problems
in statistical mechanics.
More recently, dynamical many-urn models have been investigated
in the mean-field geometry.
These exactly solvable models exhibit characteristic features
of non\-equilibrium dynamics
referred to as aging, such as the scaling of the correlation
and response functions in the two-time plane
and the violation of the \fd theorem.

This paper contains a didactic introduction to urn models~(Section~1),
a pre\-sen\-ta\-tion of static and dynamical properties
of many-urn models~(Section~2),
a reminder of the main characteristic features of aging~(Section~3),
and an overview of recent results on two dynamical urn models,
namely the backgammon model~(Section~4) and the zeta urn model~(Section~5).

\section{Urn models}
\subsection{The Ehrenfest urn model}

This model was devised by P and T Ehrenfest in 1907~\cite{ehr}
in their attempt to critically review Boltzmann's $H-$theorem.
It is defined as follows:
$N$ balls are distributed in two urns (or boxes).
At random times, given by a Poisson process with unit rate,
a ball is chosen at random,
and moved from the box in which it is to the other box.

Let $N_1(t)$ (resp.~$N_2(t)=N-N_1(t)$) be the number of balls
in box number~1 (resp.~number 2) at time $t$.
For any initial configuration, the system relaxes to equilibrium
for infinitely long times.
The equilibrium state is characterized by a binomial distribution
of the number~$N_1$:
\beq
f_{k,\eq}=\PP_\eq(N_1=k)=2^{-N}\bin{N}{k}\quad(k=0,\dots,N),
\label{feq}
\eeq
reminiscent of the Maxwell-Boltzmann statistics
for indistinguishable par\-tic\-les.

As noticed first by Kohlrausch and Schr\"odinger~\cite{ks},
$N_1(t)$ can be viewed as the co-ordinate of a fictitious walker.
The temporal evolution of
the occupation probabilities $f_k(t)=\PP(N_1(t)=k)$
is governed by the master equation
\beq
\frac{\d f_k(t)}{\d t}=\frac{k+1}{N}f_{k+1}(t)+\frac{N+1-k}{N}f_{k-1}(t)
-f_k(t).
\label{df}
\eeq
Indeed, a move of a ball from box number~1 to box number~2
(resp.~from box number~2 to box number~1) yields $k\to k-1$ (resp.~$k\to k+1$),
and occurs with a rate $k/N$ (resp.~$(N-k)/N$) per unit time.
The master equation~(\ref{df}) describes a non-uniformly biased
random walk over the integers $0,\dots,N$~\cite{ks,kac}.
The first (resp.~the second) term in the right-hand side
is absent for $k=N$ (resp.~$k=0$).

The spectrum of relaxation rates of the model
can be derived by looking for solutions to~(\ref{df})
of the form $f_k(t)=\phi_k\,\e^{-\lam t}$.
One thus obtains $\lam_m=2m/N$, with $m=0,\dots,N$.
The static solution $(m=0)$ is nothing but
the equilibrium distribution~(\ref{feq}).
The occupation probabilities have an exponential convergence
to their equilibrium values, with a relaxation time
\be
\tau_\eq=1/\lam_1=N/2.
\ee

There are, however, much larger time scales in the Ehrenfest model.
Consider indeed the length of time it takes for box number~1 to get empty,
i.e., the first time $t_0$ such that $N_1(t_0)=0$.
This time depends on the initial state and on the whole history of the system.
Its mean value $T_0=\mean{t_0}$ can be evaluated as follows.
The equilibrium probability for box number~1 to be empty,
$f_{0,\eq}=2^{-N}$, is exponentially small in the number of balls.
It is therefore expected that the typical time
needed to reach this very rare event scales as
\beq
T_0\approx1/f_{0,\eq}=2^N.
\label{t0}
\eeq
This result can be derived in a more rigorous way.
Related matters are discussed in References~\cite{lip,mur}.
The process of emptying one of the boxes
is therefore characterized by an exponentially large time $T_0$.
Equation~(\ref{t0}) can be recast as an Arrhenius-like~law:
\beq
T_0\sim\e^{S_0},
\label{ar}
\eeq
where $S_0=N\ln 2$ is the entropy difference
between the equilibrium state of the model,
where each box contains one half of the balls, up to fluctuations,
and the configuration where box number~1 is empty.
This entropy difference
plays the role of a reduced activation energy in~(\ref{ar}).
This is an elementary example of an entropy barrier.

\subsection{The Monkey urn model}

Let us introduce a variant of the Ehrenfest urn model,
which we choose to call the Monkey urn model,
because it corresponds to the image of a monkey playing at exchanging
balls between two boxes.
The key difference with the previous case is that now
at each time step a box (either 1 or 2) is chosen at random,
and one ball is moved from the chosen box (provided it is non-empty)
to the other one.
The choice of a box, instead of a ball, induces drastic changes
in the statics and dynamics of the model.

The equilibrium state of the Monkey urn model
is now characterized by uniform occupation probabilities
\beq
f_{k,\eq}=\PP_\eq(N_1=k)=\frac{1}{N+1}\quad(k=0,\dots,N).
\label{mfeq}
\eeq

The evolution of the occupation probabilities is governed
by the master equation
\be
\frac{\d f_k(t)}{\d t}=\frac{1}{2}\,\big(f_{k+1}(t)+f_{k-1}(t)-2f_k(t)\big),
\ee
describing a usual random walk over the integers $0,\dots,N$,
with appropriate reflecting boundary conditions.

The relaxation rates of the model read
$\lam_m=2\sin^2\big(m\pi/(2(N+1))\big)$, with $m=0,\dots,N$.
The relaxation time reads therefore
\beq
\tau_\eq=\frac{1}{\lam_1}\approx\frac{2}{\pi^2}\,N^2,
\label{mtau}
\eeq
where the scaling $\tau_\eq\sim N^2$ originates in the diffusive nature of the
motion of balls.

Because of the flatness of the equilibrium distribution~(\ref{mfeq}),
there are no entropy barriers in the Monkey urn model.
In particular, the mean time $T_0$ for a box to get empty scales as
$N^2$, proportionally to the relaxation time~(\ref{mtau}).

\section{Many-urn models: statics and dynamics}

We now turn to the discussion of static and dynamical properties
of two classes of many-urn models,
generalizing the Ehrenfest and the Monkey urn models
described so far~\cite{zeta}.
This classification encompasses examples of urn models studied in recent years.
The backgammon model~\cite{ritort} is a prototype of the Ehrenfest class,
while model B of References~\cite{gbm,dgc}
and the zeta urn model~\cite{dgc,zeta} belong to the Monkey class.

A many-urn model consists of $N$ balls, distributed among $M$ boxes.
Box number~$i$ contains $N_i$ balls, with
\be
\sum_{i=1}^M N_i=N.
\ee
Most investigations of many-urn models have dealt with the thermodynamic limit
($M\to\infty$, $N\to\infty$), at a fixed density $\rho=N/M$ of balls per box.

\subsection{Statics}

Two ingredients are necessary in order to define
static properties of many-urn models
within the framework of statistical mechanics:

\subsubsection*{A priori statistics.}

To each configuration $\ens$ is attributed an a priori statistical weight
\beq
W\big(\ens\big)=\left\{\matrix{
\displaystyle{\prod_{i=1}^M\frad{1}{N_i!}}
\quad&\hbox{(Ehrenfest class),}\hfill\cr\cr
1\hfill&\hbox{(Monkey class).}\hfill
}\right.
\label{apri}
\eeq
The formula for the Ehrenfest class generalizes the binomial law~(\ref{feq}),
and corresponds to the Maxwell-Boltzmann prescription
for indistinguishable particles.
The formula for the Monkey class
generalizes the flat distribution~(\ref{mfeq}),
and formally corresponds to Bose-Einstein statistics for the occupation
numbers of quantum states~\cite{kim}.

\subsubsection*{Hamiltonian.}

The Hamiltonian of a many-urn model is defined as the sum
of in\-de\-pen\-dent contributions of boxes, of the form
\beq
\H\big(\ens\big)=\sum_{i=1}^M E(N_i),
\label{ham}
\eeq
so that the unnormalized Boltzmann weight of a configuration $\ens$
at tem\-pe\-ra\-ture $T=1/\beta$ is
\beq
\e^{-\beta\H(\ens)}=\prod_{i=1}^M p_{N_i},
\label{bol}
\eeq
with
\beq
p_{N_i}=\e^{-\beta E(N_i)}.
\label{bolp}
\eeq

\subsubsection*{ }

The canonical partition function thus reads
\be
Z(M,N)=\sum_\ens
W\big(\ens\big)\,\e^{-\beta\H(\ens)}\;\delta\!\left(\sum_iN_i,N\right).
\ee
Using~(\ref{apri}) and~(\ref{bol}),
and a contour-integral representation of the Kronecker symbol, we are left with
\beq
Z(M,N)=\oint\frac{\d z}{2\i\pi z^{N+1}}\left[P(z)\right]^M,
\label{contour}
\eeq
where
\beq
P(z)=\sum_{k\ge0}p_kz^k\times\left\{\matrix{
1/k!\quad&\hbox{(Ehrenfest class),}\hfill\cr
1\hfill&\hbox{(Monkey class).}\hfill
}\right.
\label{pdef}
\eeq

Static properties of many-urn models are therefore entirely encoded in
the temperature-de\-pen\-dent generating series $P(z)$
of the Boltzmann weights $p_k$ of~(\ref{bolp}).
In the thermodynamic limit, at fixed density $\rho$, the free energy per box,
\be
\beta F=-\lim_{M\to\infty}\frac{1}{M}\ln Z(M,N)\quad(N\approx M\rho),
\ee
can be obtained by evaluating the
contour integral in~(\ref{contour}) by the saddle-point method.
The saddle-point value $z_s$
depends on temperature and density through the equation
\beq
\frac{z_s P'(z_s)}{P(z_s)}=\rho,
\label{col}
\eeq
and the free energy per box reads
\beq
\beta F=\rho\ln z_s-\ln P(z_s).
\label{free}
\eeq

The equilibrium occupation probabilities $f_{k,\eq}$
of box number~$i$ do not depend on the box under consideration.
(In the sequel we will refer to box number~1 for definiteness.)
We have
\beq
f_{k,\eq}=\frac{p_kz_s^k}{P(z_s)}\times\left\{\matrix{
1/k!\quad&\hbox{(Ehrenfest class),}\hfill\cr
1\hfill&\hbox{(Monkey class).}\hfill
}\right.
\label{feqdef}
\eeq

The above results further simplify at infinite temperature,
where $p_k=1$, independently of the form of the energy function.
Equations~(\ref{contour})--(\ref{feqdef}) become
\beq
\matrix{
P(z)=\e^z,\quad
Z(M,N)=\frad{M^N}{N!},\quad
z_s=\rho,\hfill\cr
\beta F=\rho\ln\rho-\rho,\quad
f_{k,\eq}=\e^{-\rho}\,\frad{\rho^k}{k!}\hfill
}
\label{infeh}
\eeq
for the Ehrenfest class,
and
\beq
\matrix{
P(z)=\frad{1}{1-z},\quad
Z(M,N)=\frad{(M+N-1)!}{N!(M-1)!},\quad
z_s=\frad{\rho}{\rho+1},\hfill\cr
\beta F=\rho\ln\rho-(\rho+1)\ln(\rho+1),\quad
f_{k,\eq}=\frad{\rho^k}{(\rho+1)^{k+1}}\hfill
}
\label{infmo}
\eeq
for the Monkey class.

\subsection{Dynamics}

An elementary step of the dynamics consists in moving a ball
from a departure box~$d$ to an arrival box~$a$.
A move is proposed to the system at every time step $\delta t=1/M$.
The precise definition of a dynamical many-urn model
should state the following points:

\subsubsection*{Geometry.}

The departure box $d$ and the arrival box $a$ have to be connected
to each other.
Most investigations of dynamical urn models
have dealt with the mean-field geometry,
where any two boxes are connected to each other.

Dynamical urn models can also be defined on regular lattices
in finite dimension, with the convention that urns are located
at the sites of the lattice, and that only nearest neighbors are connected.

\subsubsection*{A priori dynamical rule.}

The proposed moves are chosen according to some a priori dynamical rule.
There are three natural rules~\cite{gbm}:

\begin{itemize}

\item{\it Rule A (ball-box):}
Choose a ball (this defines the departure box $d$).
Choose an arrival box $a$, connected to~$d$.

\item{\it Rule B (box-box):}
Choose a departure box $d$.
Choose uniformly a ball in $d$.
Choose an arrival box $a$, connected to~$d$.

\item{\it Rule C (ball-ball):}
Choose a first ball (this defines the departure box $d$).
Choose a second ball (this defines the arrival box $a$).
Propose the move if the boxes $d$ and $a$ are connected.

\end{itemize}

The urn models described in Section~1 suggest what should be
the right re\-la\-tion\-ship between a priori statistics and dynamical rule.
It indeed turns out that Rule~A corresponds to the Ehrenfest class,
while Rule~B corresponds to the Monkey class.
Rule~C corresponds to yet another class of dynamical urn models,
with no obvious static counterpart~\cite{gbm}.

\subsubsection*{Thermal rule.}

The following thermal rules respect detailed balance:

\begin{itemize}

\item{\it Metropolis:}
A proposed move is performed with probability
\beq
\min\big(1,\exp(-\beta\Delta\H)\big),
\label{metro}
\eeq
where $\Delta\H=\H\big(\enspr\big)-\H\big(\ens\big)$ is the energy difference
between the con\-fig\-ura\-tions $\ens$ before and $\enspr$ after the proposed
move.

\item{\it Heat-bath:}
Among all the possible a priori choices for the arrival box $a$,
one is chosen with an appropriately normalized probability,
proportional to the Boltzmann weight $\exp\big(-\beta\H\big(\enspr\big)\big)$
of the configuration after the move.

\end{itemize}

\subsubsection*{ }

Throughout the following, we restrict ourselves to the mean-field geometry,
and to a homogeneous disordered initial state, such as, e.g.,
the infinite-temperature equilibrium state~(\ref{infeh}), (\ref{infmo}).

The key quantities are again the occupation probabilities
$f_k(t)=\PP(N_1(t)=k)$.
Their evolution is given by the master equation
\beq
\frac{\d f_k(t)}{\d t}=\sum_{\l\ge0}
\big(\pi_{k+1,\l}(t)+\pi_{\l,k-1}(t)-\pi_{k,\l}(t)-\pi_{\l,k}(t)\big),
\label{master1}
\eeq
where $\pi_{k,\l}(t)$ denotes the contribution of a move from the departure box
$d$, containing $N_d=k$ balls, to the arrival box $a$,
containing $N_a=\l$ balls.

The a priori dynamical rule yields
\be
\pi_{k,\l}(t)=f_k(t)f_\l(t)W_{k,\l}(t)(1-\delta_{k,0})\times
\left\{\matrix{
k\quad&\hbox{(Ehrenfest class),}\hfill\cr
1\hfill&\hbox{(Monkey class).}\hfill
}\right.
\ee
The factor $(1-\delta_{k,0})$ accounts for the fact that
an actual move only takes place if the departure box $d$ is not empty.
The acceptance rates $W_{k,\l}(t)$ depend on the thermal part of the rule.
With the notation~(\ref{bolp}), the rates for the Metropolis rule~\cite{dgc},
\be
W_{k,\l}=\min\left(1,\frac{p_{k-1}p_\l}{p_kp_{\l+1}}\right),
\ee
are independent of time,
while those for the heat-bath rule~\cite{dgc},
\be
W_{k,\l}(t)=\frac{p_{\l+1}}{p_\l}
\left(\sum_{\l\ge0} f_\l(t)\frac{p_{\l+1}}{p_\l}\right)^{-1},
\ee
only depend on the label $\l$ of the arrival box.

In analogy with~(\ref{df}),
(\ref{master1}) can be recast as the master equation of
a random walk over the positive integers $(k=0,1,\dots)$:
\beq
\left\{\matrix{
\frad{\d f_k(t)}{\d t}=\mu_{k+1}(t)f_{k+1}(t)+\lam_{k-1}(t)f_{k-1}(t)\hfill\cr
{\hskip 1.025truecm}-\big(\mu_k(t)+\lam_k(t)\big)f_k(t)
{\hskip 2.5truecm}(k\ge1),\hfill\cr
\frad{\d f_0(t)}{\d t}=\mu_1(t)f_1(t)-\lam_0(t)f_0(t),\hfill
}\right.
\label{mdf}
\eeq
where
\beq
\lam_k(t)=\frac{1}{f_k(t)}\sum_{\l\ge0}\pi_{\l,k}(t),\quad
\mu_k(t)=\frac{1}{f_k(t)}\sum_{\l\ge0}\pi_{k,\l}(t)
\label{lamu}
\eeq
are, respectively, the hopping rate to the right,
corresponding to $a=1$, i.e., $N_1=k\to N_1=k+1$,
and to the left, corresponding to $d=1$, i.e., $N_1=k\to N_1=k-1$.
The equation for $f_0(t)$ is special, as $\lam_{-1}(t)=\mu_0(t)=0$.

The master equation~(\ref{mdf}) preserves the sum rules
\beq
\ds{\sum_{k\ge0}f_k(t)=1},\quad
\ds{\sum_{k\ge0}k\,f_k(t)=\mean{N_1(t)}=\rho},
\label{sum}
\eeq
expressing the conservation of probability
and of the number of balls.

The rates $\lam_k(t)$ and $\mu_k(t)$, defined in~(\ref{lamu}),
depend on the occupation pro\-ba\-bi\-li\-ties $f_k(t)$ themselves,
so that the master equation~(\ref{mdf}) is actually a nonlinear set
of first-order equations for the $f_k(t)$.
The equilibrium occupation probabilities~(\ref{feqdef})
are recovered as their unique stationary state.

\section{Equilibrium vs.~nonequilibrium dynamics: two-time quantities}

Nonequilibrium properties of dynamical urn models
such as the back\-gam\-mon model~\cite{ritort,gbm,bfr,bgl}
or the zeta urn model~\cite{dgc,zeta}
have been extensively studied in recent years.

These models are simple enough to allow analytical investigations,
and rich enough to illustrate most features
of non\-eq\-ui\-li\-bri\-um dynamics referred to as aging.
The latter include the scaling properties of
the correlation and response functions of various observables,
and of their \fd ratios.
In this section we recall how the temporal evolution of these
two-time quantities
can be derived within the master-equation formalism of Section~2.

We consider a dynamical urn model of either class,
in the thermodynamic limit.
For definiteness, our observable will be the local density of balls,
defined as the fluctuating population $N_1(t)$ of box number~1, say.
The second of the sum rules~(\ref{sum})
ensures that the mean density $\mean{N_1(t)}=\rho$ is constant in time.
Its fluctuations $N_1(t)-\rho$ exhibits, however,
non-trivial dynamical properties, that can be characterized
by two-time correlation and response functions, to be defined below.
Hereafter, time variables will be such that $0\le s\le t$.
The earlier time $s$ is referred to as the waiting time,
the later time $t$ as the observation time.

\subsection{Correlation function}

The (local, connected) two-time correlation function
of the density is defined as
\be
C(t,s)=\mean{N_1(s)N_1(t)}-\rho^2.
\ee
This can be recast as
\be
C(t,s)=\sum_{k\ge1}k\,\g_k(t,s)-\rho^2,
\ee
where the quantities
\be
\g_k(t,s)=\sum_{j\ge1}j\,f_j(s)\,\PP\{N_1(t)=k\mid N_1(s)=j\}
\ee
assume the initial values
\be
\g_k(s,s)=k\,f_k(s)
\ee
at $t=s$.
Their temporal evolution for $t\ge s$ is given
by a master equation similar to~(\ref{mdf}):
\beq
\left\{\matrix{
\frad{\dpar\g_k(t,s)}{\dpar t}
=\mu_{k+1}(t)\g_{k+1}(t,s)+\lam_{k-1}(t)\g_{k-1}(t,s)\hfill\cr
{\hskip 1.38truecm}-\big(\mu_k(t)+\lam_k(t)\big)\g_k(t,s)
{\hskip 2.5truecm}(k\ge1),\hfill\cr
\frad{\dpar\g_0(t,s)}{\dpar t}=\mu_1(t)\g_1(t,s)-\lam_0(t)\g_0(t,s).\hfill
}\right.
\label{mdg}
\eeq
The rates $\lam_k(t)$ and $\mu_k(t)$, defined in~(\ref{lamu}),
only depend on the $f_k(t)$,
so that~(\ref{mdg}) are linear equations for the~$\g_k(t,s)$.

\subsection{Response function}

The (local) response function measures the influence on the mean density
in box number~1 of a perturbation in the canonically conjugate variable,
i.e., the local chemical potential acting on the same box.

Suppose that box number~1 is subjected to a small time-dependent
chemical potential $\alpha_1(t)$,
so that the total Hamiltonian of the system is now
\be
\H\big(\ens\big)=\sum_{i=1}^M E(N_i)+\alpha_1(t)N_1.
\ee
The mean density in box number~1 reads
\be
\mean{N_1(t)}=\rho+\int_0^t R(t,s)\,\alpha_1(s)\,\d s+\cdots,
\ee
where only the term linear in $\alpha(s)$ has been written explicitly,
and causality has been used.
The kernel of the linear response:
\be
R(t,s)=\frac{\delta\mean{N_1(t)}}{\delta\alpha_1(s)},
\ee
is called the two-time response function.
The temporal evolution of this function can again be determined
by means of the master-equation formalism.

\subsection{\Fd theorem: equilibrium vs.~nonequilibrium behavior}

Dynamical urn models, as any statistical-mechanical model,
have a fast con\-ver\-gen\-ce to equilibrium,
with a finite relaxation time $\tau_\eq$,
in their high-temperature (disordered) phase.
If the earlier time exceeds the relaxation time
($s\gg\tau_\eq$), the system is at equil\-ibri\-um.
One-time quantities assume their equilibrium values.
Two-time quan\-ti\-ties, such as the correlation and response functions,
are invariant under time translations:
\beq
C(t,s)=C_\eq(\tau),\quad R(t,s)=R_\eq(\tau),
\label{tti}
\eeq
where
\be
\tau=t-s\ge0.
\ee
Furthermore, these quantities are related by the \fd theorem
\beq
R_\eq(\tau)=-\frac{1}{T}\,\frac{\d C_\eq(\tau)}{\d\tau}.
\label{fdt}
\eeq

Some many-urn models exhibit a more interesting nonequilibrium behavior
at lower temperatures, where the relaxation time $\tau_\eq$
becomes either very large or infinite,
so that the waiting time $s$ and the observation time $t$
can be much smaller than $\tau_\eq$.

In such a nonequilibrium situation,
both time-translation invariance~(\ref{tti})
and the \fd theorem~(\ref{fdt}) are violated in general.
It has become customary~\cite{ck,ckp,rev1,rev2}
to characterize departure from equilibrium by the \fd ratio
\beq
X(t,s)=T\,\frac{R(t,s)}{\frad{\dpar C(t,s)}{\dpar s}}.
\label{xdef}
\eeq
In general, this dimensionless quantity depends on both times $s$ and $t$
and on the observable under consideration.
It often exhibits a non-trivial scaling behavior in the two-time plane,
which is viewed as one of the salient features of aging.

The definition~(\ref{xdef}) can be rephrased
in terms of an effective temperature~\cite{ckp}:
\be
R(t,s)=\frac{1}{T_\eff(t,s)}\,{\frad{\dpar C(t,s)}{\dpar s}},\quad
T_\eff(t,s)=\frac{T}{X(t,s)}.
\ee
At equilibrium, the \fd theorem is valid,
so that $X(t,s)=1$ and $T_\eff(t,s)=T$.
For a nonequilibrium situation started from a disordered initial state,
such as~(\ref{infeh}), (\ref{infmo}),
the effective temperature is expected to lie somewhere between
the formal initial temperature $T(0)=\infty$ and the final one,
$T(\infty)=T$, i.e.,
\be
0\le X(t,s)\le1.
\ee
These inequalities are indeed obeyed in all the known cases.

\section{The backgammon model}

The backgammon model is a simple example of a system
which exhibits slow relaxation due to entropy barriers~\cite{ritort}.
This mean-field many-urn model belongs to the Ehrenfest class,
with Metropolis dynamics, and the following choice of an energy function:
\be
E(N_i)=-\delta(N_i,0).
\ee
The Hamiltonian~(\ref{ham}) is therefore equal
to minus the number of empty boxes,
so that balls tend to condensate in fewer and fewer boxes as times passes,
at least at low temperature.

\subsection{Statics}

The equilibrium occupation probabilities read
\be
f_{k,\eq}=\e^{-\Leq}\frac{\Leq^k}{k!}\times
\left\{\matrix{
1\hfill&(k\ge1),\hfill\cr
\e^{\beta}\quad&(k=0).\hfill
}\right.
\ee
The parameter $\Leq$, equal to $z_s$ in the formalism of Section~2,
represents the mean population of a non-empty box.
It is related to temperature and density by
\be
\e^{\beta}=1+\left(\frac{\Leq}{\rho}-1\right)\,\e^{\Leq},
\ee
implying
\be
\Leq\approx\beta-\ln\frac{\beta}{\rho}
\ee
at low temperature.
The roughly linear divergence of $\Leq$
is the signature of a zero-temperature condensation transition.
The density plays no particular role in this phase transition,
so that we set $\rho=1$ in the sequel.

The slow energy relaxation at low temperature can be understood,
at least qualitatively, in terms of the concept of entropy barrier,
introduced for the Ehrenfest model.
The late stages of the dynamics indeed consist in emptying boxes
whose typical population is $\Leq$.
The formula~(\ref{t0}) suggests the scaling law $\ln\tau_\eq\sim\Leq\sim\beta$.
The relaxation time of the energy is indeed known~\cite{bgl} to read
\beq
\matrix{
\tau_\eq\approx\frad{\Leq-1}{\Leq}\,I(\Leq)
\approx\frad{\e^\Leq}{\Leq}\left(1+\frad{1}{\Leq^2}+\cdots\right)\hfill\cr
{\hskip 4.4truemm}
\approx\frad{\e^\beta}{\beta^2}\left(1+\frad{2\ln\beta+1}{\beta}+\cdots\right)
\hfill
}
\label{tl}
\eeq
at low temperature,
with the other characteristic times remaining of order $\Leq$.
The function
\be
I(\L)=\int_0^1\d z\,\frac{\e^{\L z}-1}{z}={\rm Ei}(\L)-\ln\L-\g,
\ee
where ${\rm Ei}$ is the exponential integral and $\g$ is Euler's constant,
plays a central role in the analysis of the low-temperature regime.

\subsection{Nonequilibrium dynamics: logarithmic relaxation}

In the present case,
the master equation~(\ref{mdf}) for the occupation probabilities reads
\beq
{\hskip -0.6truecm}\left\{\matrix{
\frad{\d f_k(t)}{\d t}=\frad{k+1}{\L(t)}f_{k+1}(t)+f_{k-1}(t)
-\left(1+\frad{k}{\L(t)}\right)f_k(t)\quad(k\ge2),\hfill\cr\cr
\frad{\d f_1(t)}{\d t}=\frad{2}{\L(t)}f_2(t)+\mu(t)f_0(t)-2f_1(t),\hfill\cr\cr
\frad{\d f_0(t)}{\d t}=f_1(t)-\mu(t)f_0(t),\hfill
}\right.
\label{bmas}
\eeq
with
\be
\frac{1}{\L(t)}=1-(1-\e^{-\beta})f_0(t),\quad
\mu(t)=\e^{-\beta}+(1-\e^{-\beta})f_1(t).
\ee

The parameter $\L(t)$ again represents the mean population of a non-empty box
in the low-temperature regime, while the energy density reads
\be
E(t)=-f_0(t)\approx-1+\frac{1}{\L(t)}.
\ee

The nonequilibrium regime of interest,
called the alpha regime in the context of the physics of glasses,
corresponds to low temperature and long times.
It is natural to investigate the slow relaxation dynamics
in this regime by means of an adiabatic decoupling,
similar to the Born-Oppenheimer scheme,
between the slow evolution of $\L(t)$, due to emptying new boxes,
generalizing the time scale $\tau_\eq$,
and the fast evolution of the other degrees of freedom,
corresponding to the diffusive motion of balls between non-empty boxes,
generalizing the relaxation times of order~$\Leq$.

This decoupling scheme can be implemented in several ways.
An analytical approach, extensively used in~\cite{bgl},
is based on the asymptotic analysis of formal integral representations
of the solution to the master equation~(\ref{bmas}).
It amounts to neglecting only exponentially small terms,
either in $\L$ or in $\Leq$.
This approach shows that $\L(t)$ follows an effective Markovian dynamics
\beq
\frac{\d\L}{\d t}\approx\frac{1}{I(\L(t))}
\left(1-\frac{(\L(t)-1)\e^{\L(t)}}{(\Leq-1)\e^\Leq}\right)
\label{dl}
\eeq
throughout the alpha regime ($\L(t)$ and $\Leq$ both large).

At zero temperature, the large parenthesis in~(\ref{dl}) equals unity,
so that
\be
\L(t)\approx\ln t+\ln\ln t+\frac{\ln\ln t-2}{\ln t}+\cdots
\ee
The same logarithmic law is present at finite temperature
in the aging regime, namely for $\L(t)\ll\Leq$, i.e., $t\ll\tau_\eq$.
It is the nonequilibrium counterpart
of the equilibrium relationship~(\ref{tl}) between $\tau_\eq$ and $\Leq$.

\subsection{Nonequilibrium dynamics: two-time quantities}

The two-time quantities defined in Section~3 have been studied
for both local energy and density fluctuations,
by means of the above adiabatic approach~\cite{bgl}.
The main results are as follows.

Throughout the alpha regime,
the two-time correlation and response functions
exhibit a leading power-law scaling,
with prefactors which depend logarithmically on both times $s$ and~$t$.
Consider for definiteness zero-temperature dynamics.
The energy and density correlation functions read
\be
C_E(t,s)\approx\frac{\Phi(s)}{\Phi(t)},\quad
C_D(t,s)\approx(\L(s)-1)\frac{\L(t)\Phi(s)}{\L(s)\Phi(t)},
\ee
with
\be
\matrix{
\Phi(t)=\exp\ds{\left(\int_1^{\L(t)}\frad{\d\L}{\L}
\,\frad{\L^2\e^{-\L}I(\L)+1}{\L^2\e^{-\L}I(\L)+1-\L}\right)}\hfill\cr
{\hskip 6.55truemm}=\sqrt{t}\,\,\ln t\left(1+\frad{\ln\ln t-2}{\ln
t}+\cdots\right),\hfill
}
\ee
so that their leading long-time behavior is
\be
C_E(t,s)\approx\sqrt\frac{s\ln s}{t\ln t},\quad
C_D(t,s)\approx\sqrt\frac{s}{t}\,\ln s.
\ee

Similar scaling expressions can be obtained for the local energy
and density response functions $R_E(t,s)$ and $R_D^\pm(t,s)$.
It turns out that two density response functions have to be considered,
as a peculiarity of Metropolis dynamics,
because the acceptance rate~(\ref{metro})
is not differentiable (it has a cusp) at $\Delta\H=0$.
The associated \fd ratios are independent of the observation time $t$
throughout the alpha regime, provided $\tau=t-s\gg1$ is a macroscopic time.

For a large waiting time~$s$, one has
\be
\matrix{
{\hskip .2truemm}X_E(t,s)\approx
1-\frad{2}{\L(s)^2}-\frad{4}{\L(s)^3}-\frad{6}{\L(s)^4}+\cdots\hfill\cr
X_D^+(t,s)\approx
1-\frad{2}{\L(s)^2}-\frad{2}{\L(s)^3}-\frad{12}{\L(s)^4}+\cdots\hfill\cr
X_D^-(t,s)\approx
1-\frad{3}{\L(s)^2}-\frad{3}{\L(s)^3}-\frad{15}{\L(s)^4}+\cdots\hfill
}
\ee

The \fd theorem is thus asymptotically restored in the long-time
regime, with a leading logarithmic correction in $1/(\ln s)^2$,
which depends both on the observable (energy vs.~density fluctuations)
and on microscopic details of the model (Metropolis vs.~heat-bath dynamics).

\section{The zeta urn model}

The static zeta urn model has been introduced~\cite{bia}
as a mean-field model of discretized quantum gravity.
Its dynamics has been subsequently investigated~\cite{dgc,zeta}.
The zeta urn model belongs to the Monkey class,
with mean-field heat-bath dynamics, defined by the energy function
\beq
E(N_i)=\ln(N_i+1).
\label{ze}
\eeq

\subsection{Statics}

The logarithmic confining potential~(\ref{ze}) is such that
the Boltzmann weight
\be
p_{N_i}=(N_i+1)^{-\beta}
\ee
falls off as a power law, with a temperature-dependent exponent.
The corresponding generating series has a power-law singularity at $z_c=1$,
of the form $P_\sg(z)\sim z^{\beta-1}$, allowing a condensation transition
for finite values of temperature and density.
At low enough temperature $(\beta>2)$,
$P_\sg(z)$ vanishes more rapidly than linearly as $z\to1$,
so that there is a finite critical density, corresponding to $z_s=z_c=1$,
namely
\be
\rho_c=\frac{P'(1)}{P(1)}
=\frac{\zeta(\beta-1)-\zeta(\beta)}{\zeta(\beta)},
\ee
where $\zeta$ denotes Riemann's zeta function,
which also enters~(\ref{fcrit}),
hence the name of the model, proposed in~\cite{zeta}.
The static phase diagram of the model is as follows~\cite{bia}:

\subsubsection*{Fluid phase $(\rho<\rho_c)$.}

In this low-density phase we have $z_s<1$, so that the equilibrium
occupation probabilities~(\ref{feqdef}) decay exponentially.
The density goes continuously to its critical value
($\rho\to\rho_c^-$) as $z_s\to1^-$.

\subsubsection*{Criticality $(\rho=\rho_c)$.}

Along the critical line in the temperature-density plane,
the equilibrium occupation probabilities
\beq
f_{k,\eq}=\frac{(k+1)^{-\beta}}{\zeta(\beta)}
\label{fcrit}
\eeq
obey a zeta distribution, with a temperature-dependent exponent.
More ge\-ne\-ral\-ly, critical exponents and other universal quantities
depend continuously on temperature along the critical line.
The variance of the population of box number~1:
\beq
C_\eq
=\sum_{k\ge0}k^2\,f_{k,\eq}-\rho_c^2
=\frac{\zeta(\beta)\zeta(\beta-2)-\zeta(\beta-1)^2}{\zeta(\beta)^2},
\label{eqvar}
\eeq
is finite for $\beta>3$ (regular critical regime),
while it is infinite in the strong-fluctuation critical regime~($2<\beta<3$).

\subsubsection*{Condensed phase $(\rho>\rho_c)$.}

A macroscopic condensate appears in this high-density phase.
Indeed, the static configurations are such that
the critical occupation pro\-ba\-bi\-li\-ties~(\ref{fcrit})
still hold for all boxes but one,
in which an extensive number of balls,
of order $N-M\rho_c=M(\rho-\rho_c)$, is condensed.

\subsection{Nonequilibrium critical dynamics $(\rho=\rho_c)$}

Nonequilibrium properties of the zeta urn model have been
studied recently~\cite{zeta},
both at criticality and in the condensed phase,
pursuing earlier investigations~\cite{dgc}.
The most salient results are as follows.

Consider the situation of regular criticality ($\beta>3$),
with a disordered initial state such as~(\ref{infmo}).
For any large but finite time~$t$, the system looks critical, i.e.,
the occupation probabilities $f_k(t)$ have essentially converged toward
their equilibrium values~(\ref{fcrit}), for $k\ll t^{1/2}$,
while for $k\gg t^{1/2}$ the system still looks disordered,
i.e., the $f_k(t)$ fall off very fast.
At a quantitative level, this crossover behavior is described
by the scaling law
\beq
f_k(t)\approx f_{k,\eq}\,F(u),\quad u=k\,t^{-1/2}.
\label{cfsca}
\eeq
The scaling function $F(u)$ is determined
by explicitly solving the differential equation describing
the continuum limit of the master equation~(\ref{mdf}).
Among other con\-se\-quen\-ces, the variance of the population of box number~1
is found to converge to its equilibrium value~(\ref{eqvar}) as a power law:
\beq
C(t,t)=\mean{N_1(t)^2}-\rho_c^2\approx
C_\eq-\frac{2^{3-\beta}\,t^{-(\beta-3)/2}}
{(\beta-3)\,\Gamma\left((\beta+1)/2\right)\zeta(\beta)}.
\label{cvar}
\eeq

The two-time local density correlation and response functions
have been in\-ves\-ti\-ga\-ted along the same lines~\cite{zeta}.
In the nonequilibrium scaling regime ($s,t\gg1$),
these two-time quantities are found to scale as
\beq
\matrix{
C(t,s)\approx s^{-(\beta-3)/2}\,\Phi(x),\hfill\cr
{\hskip -3.05truemm}
\frad{\dpar C(t,s)}{\dpar s}\approx s^{-(\beta-1)/2}\,\Phi_1(x),\hfill\cr
R(t,s)\approx s^{-(\beta-1)/2}\,\Phi_2(x),\hfill
}
\label{csca}
\eeq
where the dimensionless time ratio
\be
x=t/s\ge1
\ee
is the temporal analogue of an aspect ratio.
As a consequence, the \fd ratio $X(t,s)$ only depends on $x$
in the scaling regime:
\be
X(t,s)\approx\X(x)=T\,\frac{\Phi_2(x)}{\Phi_1(x)}.
\ee

The above results illustrate general predictions
on nonequilibrium critical dy\-na\-mics~\cite{janssen,glglau,glcrit,glproc}.
The exponent of the waiting time~$s$ in the first line of~(\ref{csca})
already appears in~(\ref{cvar}).
It is related to the anomalous dimension of the observable under consideration,
and would read $(d-2+\eta)/z_c$ for a $d$-dimensional ferromagnet,
with $\eta$ being the equilibrium correlation exponent
and $z_c$ the dynamical critical exponent.
The scaling functions $\Phi(x)$, $\Phi_{1,2}(x)$
are universal up to an overall multiplicative constant,
and they obey a common power-law fall-off in $x^{-\beta/2}$.
The latter exponent is not related to exponents pertaining to
usual (equilibrium) critical dynamics.
It would read $-\lambda_c/z_c=\Theta_c-d/z_c$ for a ferromagnet,
where $\lambda_c$ is the critical autocorrelation exponent~\cite{huse}
and $\Theta_c$ is the critical initial-slip exponent~\cite{janssen}.

As a consequence, the dimensionless scaling function $\X(x)$ is universal,
and it admits a non-trivial limit value in the regime where both
time variables are well separated in the scaling regime:
\be
X_\infty=\lim_{s\to\infty}\lim_{t\to\infty}X(t,s)=\X(\infty).
\ee

Explicit expressions for the above scaling functions have been derived,
using a spectral decomposition in Laguerre polynomials.
Let us mention that the limit \fd ratio,
\be
X_\infty=\frac{\beta+1}{\beta+2}\quad(\beta>3),
\ee
lies in an unusually high range ($4/5<X_\infty<1$) for a critical system.
Indeed, statistical-mechanical models such as ferromagnets
are observed to have $0<X_\infty\le1/2$ at their critical point.
The upper bound $X_\infty=1/2$, corresponding to the mean-field
situation~\cite{glcrit},
is also observed in a range of simpler models~\cite{ck,rev1,rev2,glglau}.

\subsection{Nonequilibrium condensation dynamics $(\rho>\rho_c)$}

Consider now a disordered initial condition such as~(\ref{infmo})
with $\rho>\rho_c$, so that the system will build up a macroscopic condensate.
The main results concerning this situation are as follows~\cite{zeta}.

The growth of the condensate exhibits a scaling regime
for $1\ll t\ll M^2$, while finite-size effects become dominant for later times.
In this scaling regime,
the occupation probabilities $f_k(t)$ have essentially converged toward
their equilibrium values~(\ref{fcrit}) for $k\ll t^{1/2}$,
while each box has a small probability, of order~$t^{-1/2}$,
to be part of the growing condensate, and to contain a large number of balls,
of order~$t^{1/2}$.
This picture can again be turned to a scaling law, which is of the form
\be
f_k(t)\approx\frac{F(u)}{t},\quad u=k\,t^{-1/2}.
\ee
Contrary to the scaling function $F(u)$ of~(\ref{cfsca}),
neither the present scaling function nor those
entering~(\ref{cosca}) can be expressed in closed form,
because the continuum limit of the master equation~(\ref{mdf})
in the condensed phase involves a biconfluent Heun differential equation,
whose solutions are not known explicitly.

The investigation of the scaling regime of condensation dynamics
can be extended to two-time quantities~\cite{zeta}, which are found to scale as
\beq
\matrix{
C(t,s)\approx(\rho-\rho_c)s^{1/2}\,\Phi(x),\hfill\cr
{\hskip -3.truemm}
\frad{\dpar C(t,s)}{\dpar s}\approx(\rho-\rho_c)s^{-1/2}\,\Phi_1(x),\hfill\cr
R(t,s)\approx(\rho-\rho_c)s^{-1/2}\,\Phi_2(x),\hfill\cr
{\hskip -.5truemm}X(t,s)\approx\X(x)=T\,\frad{\Phi_2(x)}{\Phi_1(x)}.\hfill
}
\label{cosca}
\eeq

This condensation dynamics exhibits many peculiarities
with respect to usual phase-ordering dynamics~\cite{bray},
which takes place, e.g., if a ferromagnet
is quenched below its critical temperature.
In the latter case, domain growth and phase separation
take place in a statistically homogeneous way, at least for an infinite system.
To the contrary, in the present situation,
condensation takes place in a very inhomogeneous fashion,
as fewer and fewer boxes are involved.

Among unusual features of two-time quantities,
let us emphasize that the scaling functions $\Phi(x)$, $\Phi_{1,2}(x)$
assume finite values, both at coinciding times $(x=1)$ and in the limit
of large time separations $(x=\infty)$.
Furthermore, the limit \fd ratio $X_\infty=\X(\infty)$
depends continuously on temperature throughout the condensed phase
($\beta>2$, i.e., $T<1/2$), and it vanishes only as
\be
X_\infty=T^{1/2}-\frac{T^{3/4}}{4}+\cdots
\ee
at low temperature, while coarsening systems are known~\cite{xzero}
to have identically $X_\infty=0$ throughout their low-temperature phase.

\end{document}